\documentclass{article}
\usepackage{PRIMEarxiv}

\usepackage[utf8]{inputenc} 
\usepackage[T1]{fontenc}    
\usepackage{hyperref}       
\usepackage{url}            
\usepackage{booktabs}       
\usepackage{amsfonts}       
\usepackage{nicefrac}       
\usepackage{microtype}      
\usepackage{lipsum}
\usepackage{fancyhdr}       
\usepackage{graphicx}       
\graphicspath{{media/}}  
\usepackage{float}

\title{Shannon entropy: an econophysical approach to cryptocurrency portfolios
}

\author{
  Noé Rodriguez-Rodriguez \\
  Facultad de Ciencias \\
  Universidad Nacional Autónoma de México\\
  Ciudad de México\\
  \texttt{noe.rdz@ciencias.unam.mx} \\
   \And
  Octavio Miramontes \\
  Departamento de Sistemas Complejos \\
  Instituto de Física\\
  Universidad Nacional Autónoma de México\\
  Ciudad de México\\
  \texttt{octavio@fisica.unam.mx} \\
}

\begin{document}
\maketitle

\begin{abstract}
Cryptocurrency markets have attracted many interest for global investors because of their novelty, wide online availability, increasing capitalization and potential profits. In the econophysics tradition we show that many of the most available cryptocurrencies have return statistics that do not follow Gaussian distributions but heavy--tailed distributions instead. Entropy measures are also applied showing that portfolio diversification is a reasonable practice for decreasing return uncertainty.
\end{abstract}


\section{Introduction}
Financial mathematics and financial economics are two scientific disciplines that together are the backbone of modern financial theory.  Both disciplines make extensive use of their own models and theories coming from mathematics and traditional economics and are aimed mostly to the predictive analysis of markets. By the other hand, physics historically had an important influence on economical theory when the formalism of thermal equilibrium was inspirational in the development of the theory of economic general equilibrium\cite{smith2008classical}. In recent years there is a renewed interest in the view that economic phenomena share many aspects of physical systems and so they are susceptible for being studied under the science of complex systems\cite{anderson1988economy,arthur2014complexity, rosser2021foundations} and statistical physics\cite{stanley1996anomalous}, given rise to novel research fields such as econophysics\cite{mantegna1999introduction}.\\    

One of the first successful econophysical approaches was the discovery that stock market fluctuations are not Gaussian but heavy--tailed. Mandelbrot came to this conclusion when investigating cotton prices\cite{mandelbrot1967variation} and later Mantegna when characterizing the Milan stock market\cite{mantegna1991levy}. This discovery forced a serious rethinking that the view introduced by Louis Bachelier in the early XX Century stating that price variations are random and statistically independent\cite{bachelier2011louis}, while historically important, is not entirely appropriate. These prices are the outcome of the concurrent non-linear action of many economic agents\cite{bonanno2001levels} and so the fluctuations are usually correlated, meaning that the process is not a random Brownian walk at all. In this article we explore fluctuations in the cryptocurrency market and show for the first time that all of the most common coins analyzed fail a Shapiro–Wilk normality test and are best explained by heavy--tailed statistical models.\\

By the other hand, the entropy is a fundamental key quantity in physics both in thermodynamics and information theory. It is related to disorder and diversity and has been previously used in finances, specifically in portfolio selection theory\cite{dionisio2006econophysics}. Acceptance of entropy in economic theory has historically some reticence, for instance the one expressed by Paul Samuelson, a seminal figure in XX Century economics\cite{samuelson1990}. However it is gaining wide acceptance lately\cite{lauster1995new, jakimowicz2020role} and specially with the development of non-equilibrium thermodynamics and complex systems theory\cite{raine2006new, prigogine20052, ormos2014entropy, almog2019structural}. It is well known that portfolio diversification is a good strategy to minimize specific risks and so we will explore the use of entropy in the cryptocurrency market as a measure of return uncertainty and risk minimization \cite{dionisio2006econophysics, goetzmann2014modern, ormos2014entropy}.

\section{Methods, data and analysis}
\label{sec:headings}

In this article we used the historical daily prices of 18 cryptocurrencies that are presented in CryptoDataDowload \cite{cryptodowload} spanning at least 3 years. These are: Basic Attention Token (BAT), Bitcoin Cash (BCH), Bitcoin (BTC), Dai (DAI), Eidoo (EDO), Eos (EOS), Ethereum Classic (ETC), Ethereum (ETH), Metaverse ETP (ETP), Litecoin (LTC), Neo (NEO), OMG Network (OMG), Tron (TRX), Stellar (XLM), Monero (XMR), Verge (XVG), Ripple (XRP), and Zcash (ZEC)\cite {osterrieder2016statistics, liang2018correlation}.\\

The opening and closing prices of the cryptocurrencies, quoted in dollars, from $10/16/2018$ to $12/31/2021$ were used, giving a total of 1172 observations per cryptocurrency, to calculate daily returns. First, the distributions of the daily returns of each cryptocurrency were statistically characterized through normality tests and parametric adjustments of heavy--tailed distributions. Subsequently, an analysis where entropy functions are derived similarly as in Dionisio et al. (2006), \cite{dionisio2006econophysics}, Ormos and Zibriezky (2014) \cite{ormos2014entropy} and Mahmoud and Naoui (2017) \cite{mahmoud2017measuring}. From the set of 18 cryptocurrencies, the assets were randomly selected to compose investment portfolios, where the only premise used was that the proportion invested in each asset is $\frac{1}{N}$, being $N$ the number of assets in the portfolio. To compare entropy to standard deviation consistently, normal entropy was used since it is a function of variance.\\

In the following subsections we will review some useful concepts regarding entropy functions: the discrete entropy function, the continuous entropy function, the entropy as a measure of uncertainty, the comparison between entropy and variance and investment portfolios.

\subsection{Discrete entropy function}
Let $X$ be a discrete random variable, $\{A_1,A_2,A_3,...A_n\}$ be a set of possible events and the corresponding probabilities $p_X(x_i) = Pr(X = A_i)$, with $ p_X(x_i) \geq 0$ and $\sum_{i=1}^{n} p_X(x_i) =1$. The generalized discrete entropy function or Rényi entropy \cite{renyi1961measures, ormos2014entropy} for the variable $X$ is defined as

\begin{equation}
     H_{\alpha}(X) = \frac{1}{1-\alpha} log \left( \sum_{i=1}^{n} p_X(x_i)^{\alpha} \right) \label{eqdisc}
\end{equation}

where $\alpha$ is the order of the entropy and $\alpha \geq 0$. This order can be considered as a bias parameter, where $\alpha \leq 1$ favors rare events and where $\alpha \geq 1$ favors common events \cite{tsallis2003nonextensive}. The base of the logarithm is 2.\\

When $\alpha = 1$ is a special case of the generalized entropy that assumes ergodicity and independence, which the generalized case does not. However, substituting into \ref{eqdisc} results in division by zero. By means of L'Hôpital's rule, it can be shown that when $\alpha$ tends to 1, we have the Shannon entropy

\begin{equation}
    H_{1}(X) = - \sum_{i=1}^{n} p_X(x_i) log(p_X(x_i)) \label{discshannon}
\end{equation}
Shannon entropy produces exponential equilibrium distributions, while generalized entropy produces power law distributions.\\

\subsection{Continuous entropy function}
Let $X$ be a continuous random variable that takes values of $\mathbb{R}$ and $p_X(x)$ be the density function of the random variable. Continuous entropy is defined as
\begin{equation}
     H_{\alpha} (X) = \frac{1}{1-\alpha} ln \int p_X(x)^{\alpha} \label{eqcont}
\end{equation}

Note that the logarithm base of \ref{eqdisc} and \ref{eqcont} are different. Although the entropy depends on the base, it can be shown that the value of the entropy changes only by a constant coefficient for different bases.\\

When $\alpha = 1$, we have the Shannon entropy for the continuous case

\begin{equation}
     H_1(X) = - \int_{}^{} p_X(x) ln(p_X(x)) dx \label{eq1}
\end{equation}

The properties of discrete and differential entropy are similar. The differences are that the discrete entropy is invariant under variable changes and the continuous entropy is not necessarily so, furthermore the continuous entropy can take negative values.\\

\subsection{Entropy as a measure of uncertainty}

According to Shannon (1948) \cite{shannon1948mathematical}, an uncertainty measure $H(p_X(x_1),p_X(x_2),...,p_X(x_n))$ must satisfy:

\begin{enumerate}
\item $H$ must be continuous on $p_X(x_i)$, with $i=1,...,n$.
\item If $p_X(x_i)=\frac{1}{n}$, $H$ must be monotone increasing as a function of $n$.
\item If an option is split into two successive options, the original $H$ must be the weighted sum of the individual values of $H$.
\end{enumerate}

Shannon showed that a measure that satisfies all these properties is \ref{discshannon} multiplied by any positive constant (the constant just sets the unit of measure). Among the properties that make it a good uncertainty choice are

\begin{enumerate}
\item $H(X)=0$ if and only if all but one of $p_X(x_i)$ are zero.
\item When $p_X(x_i)=\frac{1}{n}$ i.e. when the discrete probability distribution is constant, $H(X)$ is maximum and equal to $log(n)$.
\item $H(X,Y) \leq H(X)+H(Y)$, where equality holds if and only if $X$ and $Y$ are statistically independent i.e. $p(x_i,y_j)=p(x_i)p(y_j)$.
\item Any change towards the equalization of the probabilities $p_X(x_i)$, increases $H$.
\item $H(X,Y) = H(X) + H(Y|X)$. So the uncertainty of the joint event $(Y|X)$ is the uncertainty of $X$ plus the uncertainty of $Y$ when $X$ is known.
\item $H(Y) \geq H(Y|X)$ which implies that the uncertainty of $Y$ is never increased by knowledge of $X$. Decreases, unless $X$ and $Y$ are independent, in which case it doesn't change.
\end{enumerate}

\subsection{Comparison between Entropy and Variance}
Ebrahimi et al. (1999) \cite{ebrahimi1999ordering} showed that entropy can be related to higher order moments of a distribution, thus it can offer a better characterization of $p_X(x)$ because it uses more information about the probability distribution than the variance (which is only related to the second moment of a probability distribution).\\

The entropy measures the disparity of the density $p_X(x)$ of the uniform distribution. That is, it measures uncertainty in the sense of using $p_X(x)$ instead of the uniform distribution \cite{mahmoud2017measuring}. While the variance measures an average of distances from the mean of the probability distribution. According to Ebrahimi et al. \cite{ebrahimi1999ordering}, both measures reflect concentration, but use different metrics. The variance measures the concentration around the mean and the entropy measures the density diffusion regardless of the location of the concentration. Statistically speaking, entropy is not a mean-centered measure, but takes into account the entire empirical distribution without concentrating on a specific moment. This way you can take into account the entire distribution of returns without focusing on one particular \cite{deeva2017comparing}. The discrete entropy is positive and invariant under transformations, but the variance is not. In the continuous case neither the entropy nor the variance are invariant under one-to-one transformations \cite{dionisio2006econophysics} \cite{mahmoud2017measuring}. According to Pele et al. (2017) \cite{article} entropy is strongly related to the tails of the distribution, this feature is important for distributions with heavy tails or with an infinite second-order moment, where the variance is obsolete. Furthermore, the entropy can be estimated for any distribution, without prior knowledge of its functional form. These authors found that heavy--tailed distributions generate low entropy levels, while light--tailed distributions generate high entropy values.\\

\subsection{Investment Portfolios}

A portfolio or investment portfolio is simply a collection of assets. They are characterized by the value invested in each asset. Let $w_i$ be the fraction invested in asset $i$ with $i = 1,2,..., n$, the required constraint is that

\begin{equation}
     \sum_{i=1}^{n} w_i =1
\end{equation}

We define the return $R_i$ of a common share $i$, during a certain period as

\begin{equation}
     R_i=\frac{P_{1,i}-P_{0,i}}{P_{0,i}}
\end{equation}

where $P_{0,i}$ is the price of $i$ stock at the beginning of the period and $P_{1,i}$ is the price of $i$ at the end of that period. This return is the historical return or ex post return. Whereas the total portfolio return is simply the weighted average of the expected returns of the individual securities in the portfolio

\begin{equation}
     R_T=\sum_{i=1}^{n} w_i R_i
\end{equation}

The risk of investment portfolios can be divided into specific and systematic. Systematic risk is inherent to market uncertainty, so it is not diversifiable, usually the price of an asset is affected by factors such as inflation, economic growth or economic recession, and fluctuations in the world financial market. The specific or unsystematic risk corresponds to the risk of an asset or a small group of assets due to its specific characteristics, so it is diversifiable. Forming portfolios or portfolios can reduce specific or unsystematic risk.\\

Entropy can provide similar information if we define the average mutual information between $X$ and $Y$ as

\begin{equation}
     I(X,Y) = \sum_{i=1}^{n} \sum_{j=1}^{m} p_{X,Y}(x_i,y_j) log \left( \frac{P_{X ,Y}(x_i|y_j)}{P_X(x_i)P_{Y}(y_j)} \right)
\end{equation}

for the discrete case and

\begin{equation}
     I(X,Y) = \int_{X}^{} \int_{Y}^{} p_{X,Y}(x,y) log \left( \frac{P_{X,Y}(x| y)}{P_X(x)P_{Y}(y)} \right)
\end{equation}
for the continuous case. Thus $H(X)=I(X,Y)+H(X|Y)$, where $I(X,Y)$ can be compared to systematic risk; while $H(X|Y)$ can be compared to the specific risk.\\

In practice, standard risk measures like CAPM beta or standard deviation are calculated on daily or monthly returns. We also follow this methodology. Since the returns of the values can take values of a continuous codomain, we focus mainly on the differential entropy.\\

The assumption that data and residuals follow a normal distribution is common in both portfolio management and risk analysis. Therefore, in this context, to parametrically estimate the entropy of a normal distribution we use (because information theory measures cannot be directly compared to variance in metric terms):

\begin{equation}
     NH(X) = \int p_X(x) log(\sqrt{2 \pi}\sigma) dx + \int p_X(x) \frac{(x-\bar{x})^2}{2\sigma ^2} dx = log \left( \sqrt{2 \pi e} \sigma \right)
\end{equation}

To calculate the empirical differential entropy, it is only necessary to estimate the Probability Distribution Function (PDF), for which there are essentially three methods: based on histograms, kernels and parametric (see Appendix \ref{ApA} for details).

\section{Results}
We present the results in three parts. First, it is shown that the daily returns of each cryptocurrency do not follow a normal distribution. Second, it is indicated that the daily returns have long tails. Finally, the behavior of the empirical and normal entropy is shown as a function of the number of assets in the portfolio.

\subsection{Normal distribution and Q-Q plots}

\begin{figure}[ht]
\centering
\includegraphics[width=15cm]{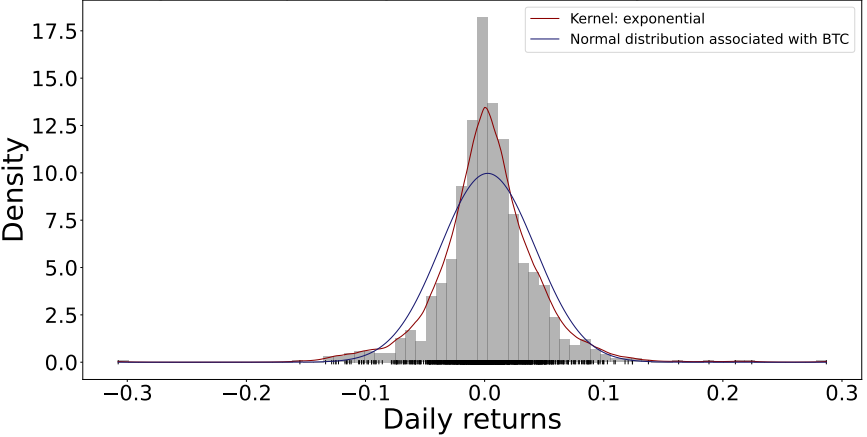}
\caption{a) Histogram of the probability distribution of the return of Bitcoin. The normal curve is shown, in blue, associated with the data set and the curve estimated using the Kernel method, in red.}
     \label{normal1}
\end{figure}

The figure \ref{normal1} shows the histogram of the probability distribution of the daily returns of Bitcoin, as an example. The blue curve shows the normal distribution with the variance and mean calculated from the empirical data. The red line shows the Kernel density estimator, the cross-validation method was used to determine the bandwidth and the Kernel that maximize the total Log-Likelihood or Log-Likelihood of the data in $X$ \cite{inbook} . Among the kernels used are: Gaussian, Epanechnikov, Tophat, exponential, linear and cosine. Bandwidth was optimized for values between 0 and 1. For all cryptocurrencies, it was found that there is a significant deviation of the distribution calculated with Kernels with respect to the normal distribution.

Quantiles are values that divide a probability distribution into equal intervals, each interval having the same fraction of the total population. Q-Q plots are commonly used to visualize data and comparatively find the type of probability distribution to which a random variable can belong, for example, if they are Gaussian, uniform, Paretian, exponential distributions, etc.
To build a Q-Q plot, the quantiles of the base distribution or the $"$theoretical quantiles$"$ are plotted on the $X$ axis and the sample quantiles are plotted on the $Y$ axis. If both sets of quantiles come from the same distribution, you should see that the points form a more or less straight line.\\

\begin{figure}[ht]
\centering
\includegraphics[width = 10cm]{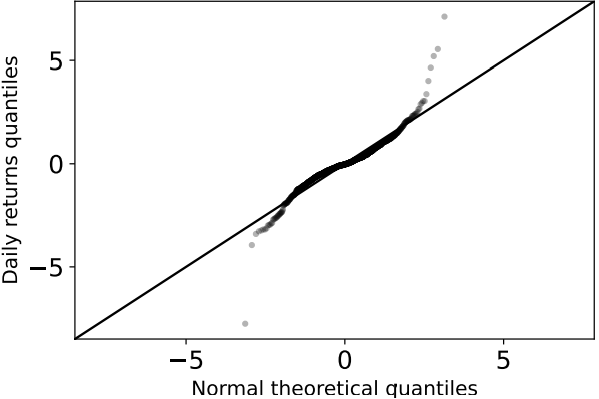}
\caption{Normal Q-Q chart of daily Bitcoin performance. The theoretical normal quantiles are plotted against the observed quantiles.}
\label{figqqplot}
\end{figure}

The figure \ref{figqqplot} shows the normal Q-Q plot of the empirical daily returns of Bitcoin. Across all cryptocurrencies, the normal distribution was found to capture the middles of the data well, but not the tails. In fact, the characteristic behavior of heavy--tailed distributions was clearly seen by the $"$S$"$ shape on the graph. This means that, compared to a normal distribution, there is more data located at the extremes of the distribution and less data at the center of the distribution.\\

The Shapiro-Wilks test posits the null hypothesis that a sample comes from a normal distribution. When applying this statistical test to the daily returns of all the cryptocurrencies studied, it was found that the $p$-value is at most of the order of $10^{-18}$. It is established as common practice that if the $p$ value is less than 0.05, the hypothesis that the sample is normal should be rejected. In our case the value $p$ is very small (almost negligible indeed) for all cryptocurrencies, which strongly indicates that the null hypothesis that daily returns follow a normal distribution \cite{shaphiro1965analysis} should be rejected. Therefore, we accept the alternative hypothesis that the returns do not present normality.\\

\subsection{Heavy--tail distributions}

Nadarajah et al. (2015) \cite{nadarajah2015note} made adjustments to the exchange rate of various fiat currencies\footnote{money that is not backed by securities or physical assets such as gold, but is backed by the government that issues it.} using flexible distributions . They used distributions such as Student-t, skewed Student-t, hyperbolic, generalized hyperbolic, generalized lambda, Skew-T, and Gaussian inverse normal to the data. According to the previous section, we must limit ourselves to the study of distributions with long tails. Osterrieder (2016) \cite{osterrieder2016statistics} performed heavy--tailed fits to different cryptocurrencies with the Student-t, Generalized Student-t, Hyperbolic, Generalized Hyperbolic, Gaussian Inverse Normal, and Asymmetric Gamma variance distribution. In both studies, they found that all the heavy--tailed distributions used gave statistically similar results, although the best fit was found with the generalized hyperbolic distribution. In addition, they concluded that the Student-t is a good option given its simplicity. Similarly, Briere et al. (2017) \cite{estadistica} reached similar results when studying parametric adjustments to the returns of the 7 most traded cryptocurrencies of 2015. Therefore, here we make adjustments to the daily returns with the t-Student distribution and the normal distribution for comparison, although it is light-tailed.\\

\begin{figure}[ht]
\centering
\includegraphics[width = 15cm]{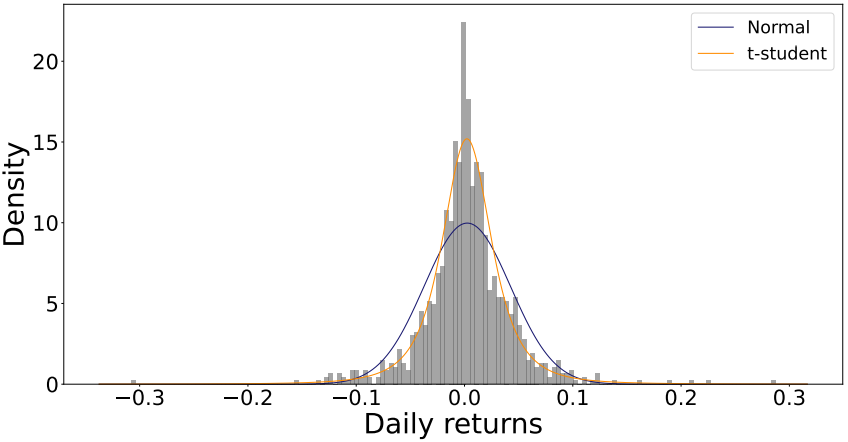}
\caption{a) Histogram of the probability distribution of the return of Bitcoin with the adjustments of a normal distribution and t-student. }
     \label{weighted1}
\end{figure}

Figure \ref{weighted1} shows the histogram of the probability distribution of Bitcoin daily performance along with the adjustments of each distribution so that the Log-Likelihood is maximized. The Log-Likelihood value for each fit is shown in the table \ref{weighed2}. Note that the Log-Likelihood is always smaller for the normal distribution while the heavy--tailed t-student distribution generates good fits.

\begin{table}[H]
 \caption{Log-likelihood of the fitted distributions.}
 \label{weighed2}
  \centering
  \begin{tabular}{lll}
    \toprule                 \\
    \cmidrule(r){1-3}
    Criptocurrency     & Normal     & t-student \\
    \midrule
BAT                                 & 1531                                                  & 1667                                                                                                                        \\ \hline
BCH                                 & 1555                                                  & 1815                                                                                                                         \\ \hline
BTC                                 & 2111                                                  & 2250                                                                                                                         \\ \hline
DAI                                 & 4395                                                  & 4871                                                                                                                         \\ \hline
EDO                                 & 1239                                                  & 1518                                                                                                                         \\ \hline
EOS                                 & 1583                                                  & 1794                                                                                                                         \\ \hline
ETC                                 & 1608                                                  & 1861                                                                                                                         \\ \hline
ETH                                 & 1831                                                  & 1915                                                                                                                        \\ \hline
ETP                                 & 1393                                                  & 1594                                                       
                                                                 \\ \hline
LTC                                 & 2111                                                  & 2250                                                                                                                         \\ \hline
NEO                                 & 1631                                                  & 1740                                                                                                                         \\ \hline
OMG                                 & 1401                                                  & 1570                                                                                                                         \\ \hline
TRX                                 & 1694                                                  & 1829                                                                                                                         \\ \hline
XLM                                 & 1590                                                  & 1808                                                                                                                         \\ \hline
XMR                                 & 1814                                                  & 1947                                                                                                                         \\ \hline
XRP                                 & 1595                                                  & 1884                                                                                                                         \\ \hline
XVG                                 & 1305                                                  & 1396                                                                                                                         \\ \hline
ZEC                                 & 1637                                                  & 1722                                                                                                                         \\ \hline
    \bottomrule
  \end{tabular}
  \label{tab:table}
\end{table}

\subsection{Portfolio Entropy}

\begin{figure}
 \includegraphics[width=16cm]{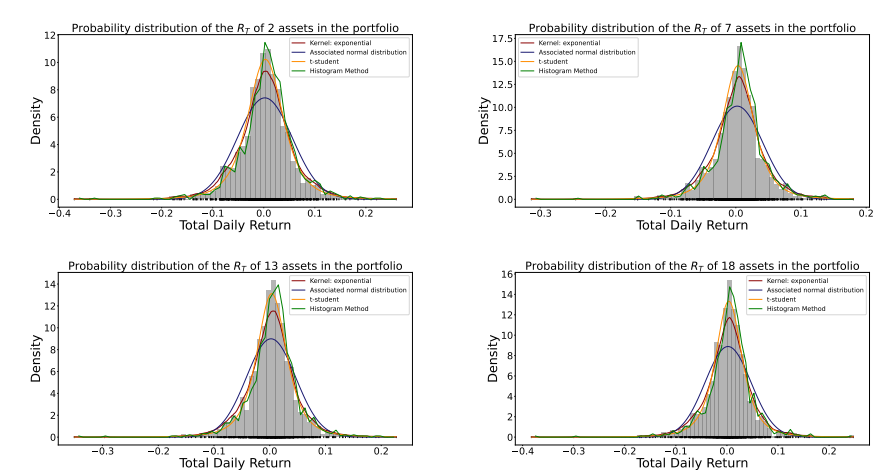}
\caption{a) Adjustment to the probability distribution of the total daily returns with $n \in \{2,7,13,18\}$ assets in the portfolio with the kernel method, parametric, histogram and the normal curve associated with the data.}
\label{totalreturn}
\end{figure}

\begin{figure}[H]
    \centering
    \includegraphics[width=5in]{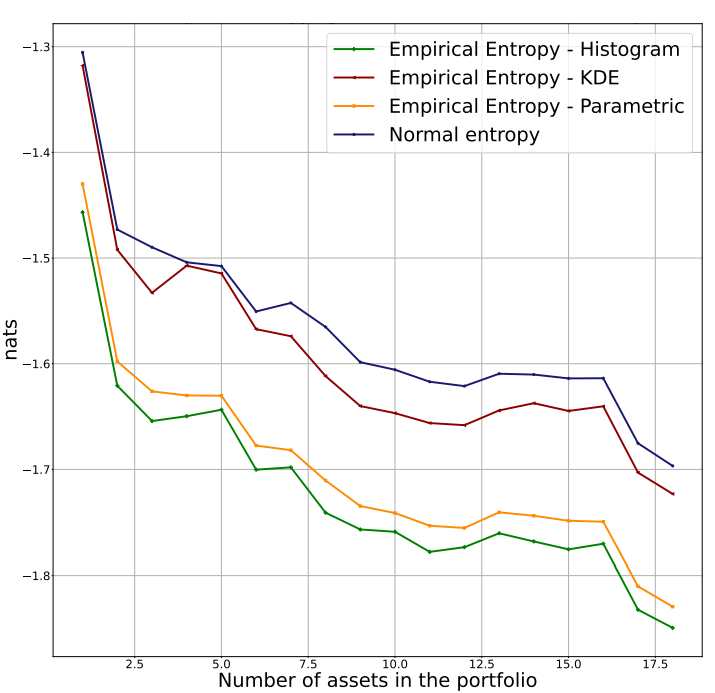}
    \caption{ Normal and empirical entropy against the number of assets in the portfolio. The decrease in entropy as the number of assets in the portfolio increases verifies the diversification effect.}
    \label{final}
\end{figure}

A portfolio was generated by adding the 18 cryptocurrencies randomly. It began by adding an asset to the portfolio, then its normal entropy and its empirical entropy were calculated using the histogram, Kernel and parametric methods. Assets continued to be added randomly until the 18 cryptocurrencies studied were added. Every time a new asset was added to the portfolio, it was necessary to calculate the normal entropy and the empirical entropy of the total return i.e., every time an asset was added, the distribution function of the total return was calculated again by means of the three methods presented. The figure \ref{totalreturn} shows the adjustments when performing the previous procedure, while the figure \ref{final} shows the normal and empirical entropy against the number of assets in the portfolio.

\section{Discussion}

In this work, some statistical properties of the yields of the studied cryptocurrencies were shown. For this, the daily returns were adjusted to different distributions such as the normal and t-student. The results show that the distributions of daily yields exhibit statistically significant leptokurtic return distributions with heavy tails. In particular, the t-Student distribution adequately describes the data and is a good option given its simplicity, which can be useful for financial risk management, where it is necessary to calculate the value at risk (VaR) and the expected deficit (ES) , but the results are also useful for investment purposes. As far as we know, this is one of the few works that investigates some statistical properties of cryptocurrencies, going beyond Bitcoin.\\

Entropy was used as a measure of uncertainty in portfolio management. Figure \ref{final} shows that the Shannon entropy behaves similarly, although not the same, as the variance, so it can serve as a measure of risk and verifies the effect of diversification since both tend to decrease when including assets in the portfolio. This is because as the number of assets in the portfolio increases, the possible number of states of the system, in this case the portfolio, progressively decreases and the uncertainty about this portfolio also decreases. We should note that in all cases the normal entropy always takes values greater than the empirical entropy, which implies that the uncertainty is less than that which would be observed if the returns were normally distributed. \\

In the study by Ormos and Zibriezky (2014) \cite{ormos2014entropy} it was found that the histogram method for calculating the Shannon entropy showed to be more efficient in terms of explanatory and predictive power and exhibited simplicity compared to the Kernel method. We observe that the fits of the figure \ref{totalreturn} to the daily total return of the generated portfolios show that the histogram method presents an overfit to the empirical data. On the other hand, the best fit is parametric followed by KDE. In these last two cases, the Log-Likelihood of the parametric method with t-Student distribution was always greater than that of the Kernel Density Estimation (KDE) method; although the Log-likelihood of both was always greater than that of the normal distribution. However, with any method, diversification is verified. This corroborates and contributes to the empirical findings made by Dionisio et al. (2006), \cite{dionisio2006econophysics}, Ormos and Zibriezky (2014) \cite{ormos2014entropy} and Mahmoud and Naoui (2017) \cite{mahmoud2017measuring} who observed the diversification of the Shannon entropy, with the difference that these authors they verified it on classical assets and did not characterize their distributions. Our results may be of use to investors, traders and portfolio managers. \\

\section{Conclusions}

We conclude that entropy observes the effect of diversification and is a more general measure of uncertainty than the standard deviation since: (i) it uses more information from the probability distribution, since it uses moments of higher order, while the deviation standard only uses the second moment; (ii) It does not depend on any particular distribution, unlike the standard deviation, which eliminates the error introduced by fitting a normal distribution to returns. This becomes evident in non-symmetric distributions with additional non-normal moments. In this way, entropy can capture the complexity of systems without the need for hypotheses that could bias the results; (iii) the entropy is independent of the mean given any distribution, so it satisfies the first-order conditions; (iv) meets the requirements that good uncertainty measures must satisfy; (v) entropy can be used for both metric and non-metric data. For this reason, it can be used as a complementary measure to traditional models based on the mean and variance that are more restrictive with assumptions that tend not to be verified empirically. However, the drawbacks of the potential use of entropy as a measure of uncertainty in portfolios should also be mentioned, one of them being that it is more complex compared to the common standard deviation. On the other hand, entropy also does not take into account the real values of the variables, so care must be taken when using it in risk analysis and portfolio selection. Also, there is always statistical bias in data measurements due to the degrees of freedom allowed in an experiment.\\

Although this is not a work on punctual prediction of the cryptocurrency market, it is illustrative to mention the relative success, in this area, of the use of nonlinear physics methods to make sustantive predictions, made among others by the ``Prediction Company'' founded by two pioneers of chaos theory in 1991, associated with the Santa Fe Institute. This is the case of Doyne Farmer and Norm Packard \cite{gwynne2001physicist}. In one of his best-known contributions \cite{farmer_SFI}, Farmer mentions the possibility of using ``adaptive dynamics'' to make predictions in natural time series and that follows the principles of neural networks or more modernly what is known as known as ``machine learning''. It basically consists of the use of systems that learn to classify patterns (output) from a time series (input). The output would ideally be an investment strategy that would aim to maximize profits and the input would be the immediate, ideally real-time, time series of market behavior. Of course, there are recent applications of machine learning to the prediction of the temporal behavior of cryptocurrencies \cite{cocco2021predictions, chen2021machine}.\\

For the temporal prediction of financial markets there are, today, more than 5,000 algorithms created by various programmers who publish them in public software libraries \cite{tradingview2020}. Today, investors use several of these algorithms in combination, although the most popular are usually based on: Average True Value (ATR), Relative Strength Index (RSI), The Moving Average Convergence/Divergence (MACD) and the Average Exponential Moving (EMA)\cite{tradingview2020}. As can be seen, the use of Moving Averages is common. They essentially help to reduce noise or fluctuations in time intervals. Some authors \cite{lambert2002modelling, maleki2020asymmetric} couple them with regression analysis to form ARMA models (Autoregressive–moving-average model) which in turn are used together with skewed (non-normal) probability distribution functions, since, as we have already confirmed here, the PDFs of the financial markets and cryptocurrencies are not normal. It is interesting to note that there are ``forecasts'' online that suggest the right times, in real time, to sell or buy cryptocurrencies and that they are mostly based on moving average models\cite{tradingview2020bitcoin}.\\

Finally, it is illustrative to mention that cryptocurrency markets have qualitative and quantitative properties similar to traditional financial markets, which allows us to speak of generic properties between the two, that is, aspects of universality, typical of complex systems.


\section{Aknowledgments}
NRR woud like to thank the Physics Institute at UNAM in Mexico City for providing financial support. Grant PRIDIF-2021.

\section{Appedix A - Probability Density Function Estimation (PDF)}\label{ApA}
Let $x_1$, $x_2$, ..., $x_n$ be the observations of the continuous random variable $X$ and $H_{\alpha,n}(X)$ be the sample estimate of $H_{\alpha}( X)$. The evaluation of the entropy is obtained by estimating the density function. The probability density function $p_X(x)$ is approximated by $p_{X,n}(x)$. Thus, the integral estimate of the entropy is

\begin{equation}
    H_{\alpha,n}(X) = \frac{1}{1-\alpha} ln \int_{A_n}^{} p_{X,n}(x)^{\alpha} dx \label{eq3}
\end{equation}

where $A_n=(min(x), max(x))$. For this reason, one of the difficulties in determining continuous entropy is that the underlying probability density function (or \textit{PDF}) is unknown. There are three methods to overcome this problem: histogram-based, kernel-based, and parametric estimators. \cite{dionisio2006econophysics, ormos2014entropy}\\

\subsubsection{Histogram}

Let $b_n=(max(x),min(x))$ be the range of the sample values; we divide the range into $k$ bins of equal width where the cuts are made at the points $t_j$. The width of each bin is $h=\frac{b_n}{k}=t_{j+1}-t_{j}$. The density function is estimated by

\begin{equation}
    p_{X_n}(x) = \frac{v_j}{nh}
\end{equation}

if $x \in (t_j, t_{j+1})$, with $v_j$ being the number of points that land in the jth bin.\\
From \ref{eq1} and \ref{eq3}

\begin{equation}
    H_{1,n}(X) = \frac{1}{n} \sum_{j=1}^{k} v_j ln \left(\frac{v_j}{nh} \right)
\end{equation}

The hyperparameter of this method is the number of bins $k$. There are several methods to choose this parameter, for example Scott's (1979) \cite{scott1979optimal} normal reference rule, Freedman-Diaconis (1981) \cite{freedman1981histogram} rule, square root rule, etc. In particular, the Freedman-Diaconis rule is quite robust and usually gives good results in practice. The method consists of minimizing the integral of the squared difference between the histogram and the density of the theoretical distribution. The bandwidth in this method is given by

\begin{equation}
    h = 2 \cdot IQR(x) \cdot n^{-1/3}
\end{equation}

where $n$ is the number of observations and $IQR(x)$ is the interquartile range of the data set. In this way the optimal number of containers is chosen as

\begin{equation}
    k = \frac{b_n \cdot n^{1/3}}{2 \cdot IQR (x)}
\end{equation}

\subsubsection{Kernel Density Estimation}

Let $x_1$, ...., $x_n$ be a sample of independent and identically distributed random variables, then an estimator of the distribution function is

\begin{equation}
    p_{X_n}(x) = \frac{1}{nh} \sum_{i=1}^{n} \int_{-\infty}^{x} K \left( \frac{u-x_i}{ h} \right) du
\end{equation}
where $K$ is the kernel function and $h$ is the bandwidth parameter. The function $K$ must be a real function such that $K(x) \geq 0$, $K(x)=K(-x)$ for all $x$ in $\mathbb{R}$, $\ int_{-\infty}^{\infty} K(x) dx=1$ and $\int_{-\infty}^{\infty}xK(x)dx=0$. The uniform convergence to the theoretical distribution has been studied regardless of the form of the kernel used \cite{article}. The hyperparameters are the Kernel and the bandwidth, the choice of Kernel determines how the influence of each observation is distributed.

\subsubsection{Parametric density estimation}

The set of probability distributions in a sample space indexed by a set $\Theta$ (parameter space) is denoted as $\mathcal{P}$. For each $\theta \in \Theta$, there is a $P_{\theta}$ member of $\mathcal{P}$ which is a distribution. Statistical models can be written as
\begin{equation}
    \mathcal{P} = \{P_{\theta} | \theta \in \theta\}
\end{equation}

These models are parametric if $\Theta \subseteq \mathbb{R}^{k} $ with $k \geq \in \mathbb{Z}^{+}$ (otherwise we would have a non-parametric model like the estimators of kernels). In which case, the model consists of continuous distributions. Parametric methods assume that the particular shape of the probability distribution function is known and only its parameters need to be estimated (in the case of a normal distribution, only its mean and variance need to be estimated).\\

\subsubsection{Maximum Likelihood Estimate}

Consider a set $(x_1,x_2,...,x_n)$ of i.i.d. Since they are identically distributed, they must have the same probability function $f(X|\theta)$. The likelihood of the data having the parameter $\theta$ is

\begin{equation}
    \mathcal{L} (\theta) = \Pi_{i=1}^{n} f(x_i|\theta)
\end{equation}
For different $\theta$'s, the probability of the data having those parameters will be different. In maximum likelihood estimation (MLE), the goal is to find the parameters $\theta$'s that maximize the likelihood function $\mathcal{L}$. The MLE assumes that:\footnote{The arguments of the maximum or $argmax$ are those points in the domain of a function at which it is maximized.}
\begin{equation}
    \hat{\theta} = \arg \max_{\theta} \left( \mathcal{L} (\theta) \right)
\end{equation}

The Log-Likelihood or Log-Likelihood function is usually used
\begin{equation}
    \mathcal{LL} (\theta) = log(\mathcal{L} (\theta)) = log \Pi_{i=1}^{n} f(x_i|\theta) = \sum_{i=1} n logf(x_i|\theta)
\end{equation}

Since the logarithm is monotonically increasing, in practice the MLE works with

\begin{equation}
    \hat{\theta} = \arg \max_{\theta} \left( \mathcal{LL} (\theta) \right)
\end{equation}

For this work, the histogram, Kernel and parametric methods were used to determine the $PDF$'s, always using the Freedman-Diaconis rule. In the last two cases, the probability density functions were calculated by maximum likelihood estimation (MLE).

\subsection[\appendixname~\thesubsection]{Parametric distributions used}

Normal distribution
\begin{equation}
p_X(x)=\frac{1}{{\sigma \sqrt {2\pi } }}e^{{{ - \left( {x - \mu } \right)^2 } \mathord{\left/ {\vphantom {{ - \left( {x - \mu } \right)^2 } {2\sigma ^2 }}} \right. \kern-\nulldelimiterspace} {2\sigma ^2 }}}
\end{equation}
for $-\infty \leq x \leq \infty$, $-\infty \leq \mu \leq \infty$ and $\sigma \geq 0$.\\

Non-standardized Student's t-distribution

\begin{equation}
     p_X(x) = \frac{K(\nu)}{\sigma} \left[ 1+\frac{(x-\mu)^2}{\sigma^2 \nu } \right]^{-( 1+\nu)^2/2}
\end{equation}
for $-\infty \leq x \leq \infty$, $-\infty \leq \mu \leq \infty$ and $\sigma \geq 0$ u $\nu \geq 0$. Where $K(\nu) = \sqrt{\nu}B(\nu/2,1/2)$ and $B$ is the beta function defined by

\begin{equation}
     B(a,b) = \int_{0}^{1} t^{a-1} (1-t)^{b-1}dt
\end{equation}

\bibliographystyle{unsrt}  
\bibliography{references}

\end{document}